# Local Area Damage Detection in Composite Structures Using Piezoelectric Transducers

Peter F. Lichtenwalner[a] and Donald A. Sofge[b]

[a]The Boeing Company, MC S1021310, P.O. Box 516, St. Louis, MO 63166
[b]NeuroDyne, Inc., One Kendall Square, Cambridge, MA 02139

## ABSTRACT

An integrated and automated smart structures approach for structural health monitoring applications is presented. This approach, called Active Damage Interrogation (ADI), utilizes an array of piezoelectric transducers attached to or embedded within the structure for both actuation and sensing. The ADI system actively interrogates the structure via broadband excitation of multiple actuators across a desired frequency range. The structure's vibration signature is then characterized by computing the transfer functions between each actuator/sensor pair, and compared to the baseline signature. Statistical analysis of the transfer function deviations from the baseline is used to detect, localize, and assess the severity of damage in the structure.

Experimental results of applying the ADI system for local area damage detection in a MD Explorer rotorcraft composite flexbeam are presented. The performance of the system in detecting and localizing internal delaminations created by low velocity impacts is quantified. The results obtained thus far indicate considerable promise for integrated structural health monitoring of aerospace vehicles, leading to the practice of condition-based maintenance and consequent reduction in vehicle life cycle costs.

**Keywords:** smart structures, health monitoring, damage detection, structural integrity, piezoelectric, vibration

## 1. INTRODUCTION

A significant area of ongoing research and development efforts in the aerospace smart structures community is the implementation of an automated structural health monitoring (SHM) system using smart sensors and actuators integrated into the structure in order to provide a "built-in-test" diagnostic and prognostic capability for the vehicle's health management system. A reliable SHM system will enable the practice of condition-based maintenance which can significantly reduce life cycle costs by eliminating unnecessary inspections, minimizing inspection time and effort, and extending the useful life of new and aging aerospace structural components. The most promising techniques under development involve the use of piezoelectric transducers to actively excite and sense the vibration characteristics of the structure and use this information to make estimates regarding the health of the structure.[1-4] These techniques have been shown to be highly sensitive in damage detection and provide several inherent advantages over "passive" approaches such as strain monitoring and acoustic emission monitoring.

This paper presents an SHM method developed at The Boeing Company-Phantom Works Division under internal research and development funding called *Active Damage Interrogation (ADI)*, (U.S. Patent Pending). The ADI system utilizes an array of piezoelectric transducers attached to or embedded within the structure for both actuation and sensing. By detecting changes in the structure's vibration signature, damage can be detected, localized, and assessed. Under previous work, the ADI system was demonstrated to be highly sensitive in detecting edge delaminations on a flexbeam test article from the MD Explorer helicopter.[5] This paper describes work conducted under funding from NeuroDyne, Inc. in support of a Small Business Innovative Research (SBIR) program sponsored by the National Science Foundation. Under this effort, the ADI system was applied for detecting and localizing internal delaminations created by low velocity impacts to the flexbeam. A description of the ADI system, experimental test setup, and experimental results are presented.

## 1.1 Overview of ADI System Operation

The Active Damage Interrogation (ADI) system is a smart structures approach for health monitoring which provides a method for detecting and localizing structural damage according to changes in the characteristic vibration signature of the structure. The ADI system uses piezoelectric (PZT) transducers attached to the structure which act as actuators for exciting the structure, and sensors for measuring the vibration response of the structure. The structure's vibration signature is then characterized by computing the transfer function between each actuator and sensor. This transfer function contains the magnitude and phase information versus frequency for the sensor response as a function of the actuator input. These transfer functions are compared to a reference, or baseline, which represents the normal "healthy" state of the structure. The baseline is generated by collecting several sets of actuator/sensor data when the structure is in a healthy state. The mean and standard deviation is then computed to produce the baseline.

The difference between the currently measured transfer function and the baseline transfer function is normalized by the baseline's standard deviation. This provides a measure of the vibration signature change in terms of the number of standard deviations from the baseline. This normalization is especially useful in compensating for varying signal-to-noise ratios (SNRs) among the sensors. The normalized difference is input to a windowed averaging process and then integrated across the frequency range to produce a cumulative average delta (CAD) which provides a single metric for indicating damage. This information processing flow is outlined graphically in the block diagram of Figure 1.

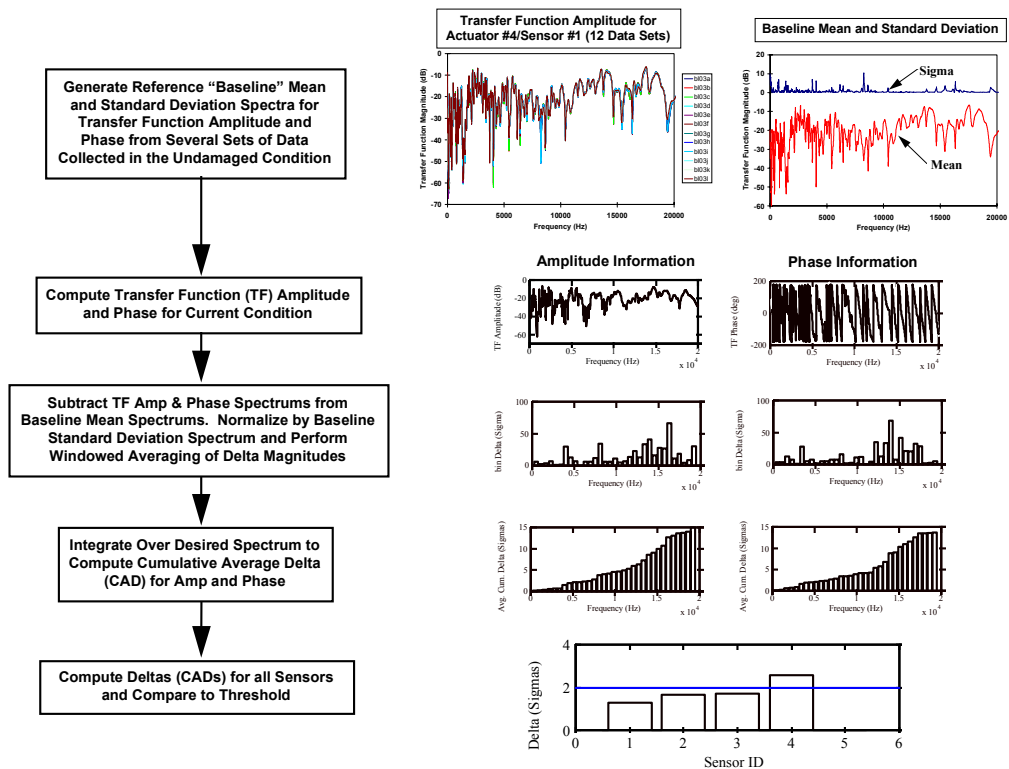

**Figure 1. ADI System Block Diagram**

The transfer function magnitude and phase CAD values computed for all sensors in relation to a single actuator are then averaged to produce a single "damage index" assigned to each piezoelectric transducer. These damage indices are then passed to downstream processing algorithms for both detection and localization of the damage. As in all statistical detection schemes, the detection threshold value must be established according to a specified cost function which optimally maximizes detection probability while minimizing the false alarm rate. This threshold should account for all unmodeled variations that normally occur in the undamaged state, and for changes due to damage smaller than the required minimum flaw size. Thus the detection threshold for the damage index must eventually be determined based on variation due to structural loading, environmental conditions, and the minimum flaw size to be detected.

## 2. EXPERIMENTAL SETUP

The MD Explorer flexbeam is a glass-epoxy composite structure which links the rotorblades to the hub and is housed inside the pitchcase as shown in Figure 2. The flexbeam is a critical load bearing structural component on the MD Explorer, since the loads from the blades are transferred to the hub and the rest of the rotorcraft through these components. Since the flexbeam is a composite structure and hence susceptible to delamination damage, it was selected as the test article for evaluating the performance of the ADI system for integrated structural health monitoring.

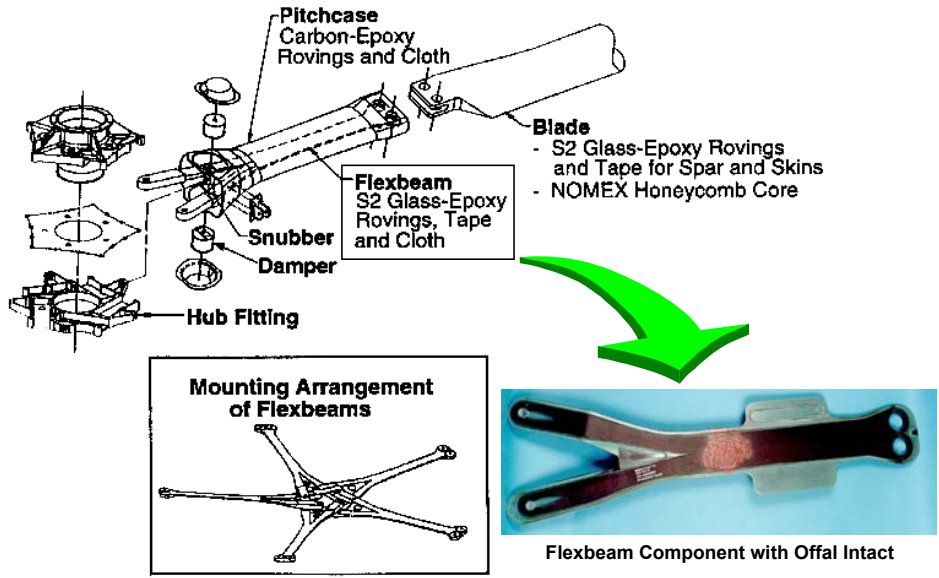

**Figure 2. Flexbeam Location on the MD Explorer**

The flexbeam test article was mounted in a cantilevered condition, as shown in Figure 3, to represent the boundary conditions of its installation within the rotor system. Four 1/2" by 1/2" cut PZT wafer transducers were attached to a localized section of the flexbeam to be used as the ADI actuators and sensors. Two positions within this localized area were specified for inducing the internal delamination damage, as shown in Figure 4.

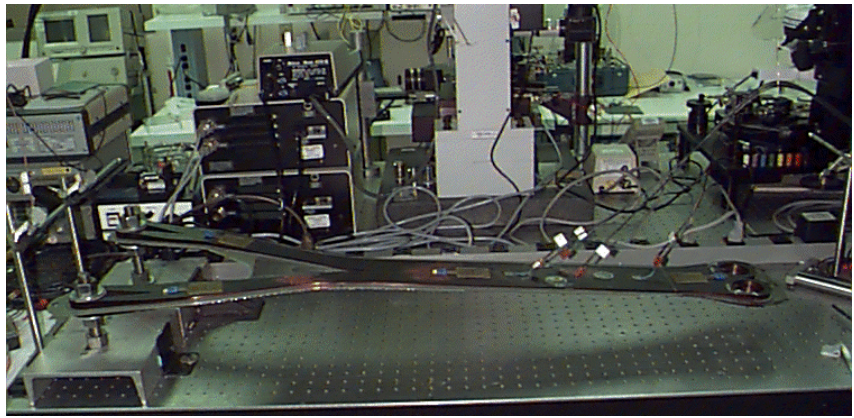

**Figure 3. MD Explorer Composite Flexbeam in a Cantilever Mount**

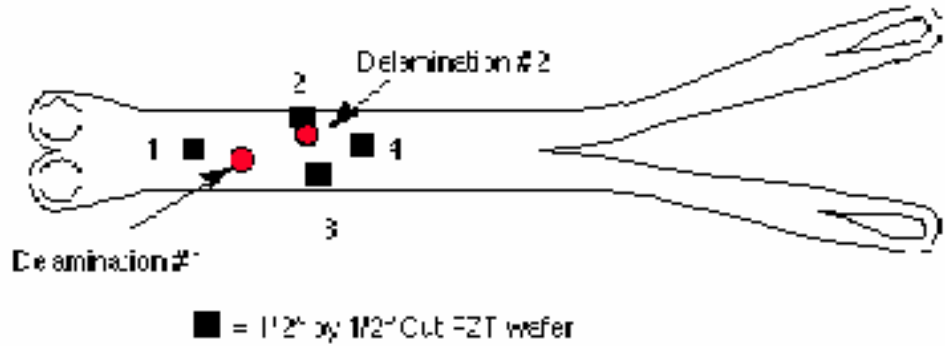

**Figure 4. Illustration of Flexbeam Showing Piezoelectric Transducers and Delamination Sites**

At each site, three delaminations of increasing size were created by successively dropping a 2.9 pound projectile with a 0.5" diameter point from heights of 1-3 feet. The impact test configuration is shown in Figure 5.

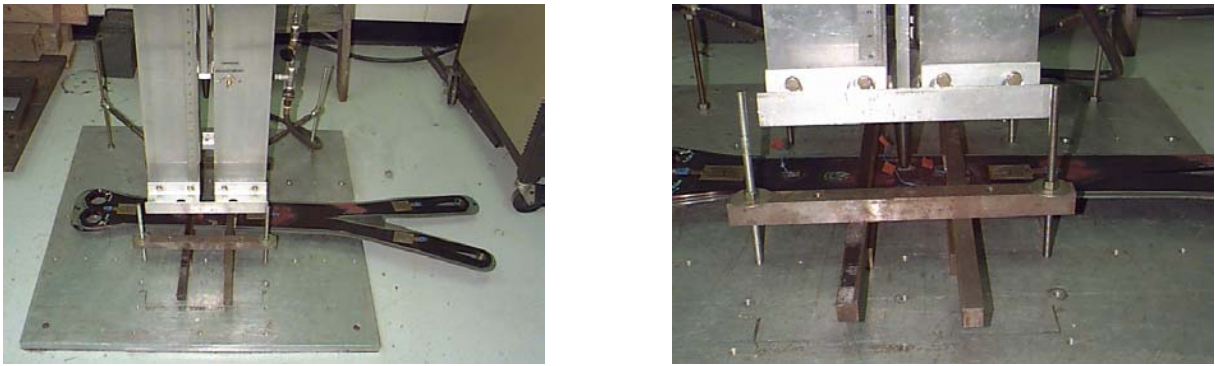

**Figure 5. Impact Setup for Creating Low Velocity Internal Delaminations**

The smallest delamination was created and an ultrasonic inspection was performed to determine the size of the delamination. This process was repeated until a desirable delamination size was achieved. Once the desirable size was achieved, the ADI system was used to collect data and evaluate the damage induced to the flexbeam. The small delamination was then enlarged to produce the medium-sized delamination. The process of ultrasonically inspecting the flexbeam, inducing damage, and collecting data was repeated three times at each damage site, giving a total of six impact delaminations with diameters ranging from 0.5 to 1.25 inches. The delaminations and PZT transducers are shown in Figure 6 along with a quarter to indicate their relative sizes.

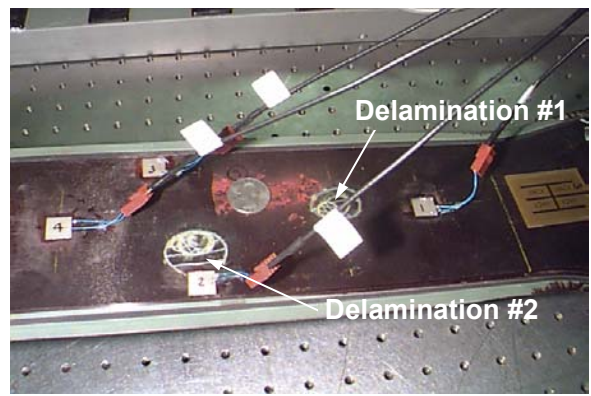

**Figure 6. Flexbeam Section with Delaminations and PZT Transducers**

## 3. EXPERIMENTAL RESULTS

Thirteen sets of data were collected with the flexbeam in the undamaged state. These were used to compute the first set of baseline transfer functions which represent the vibration characteristics of the flexbeam in the undamaged state. After the large delamination was induced at site #1, a second set of baselines were generated which included the first delamination. This second baseline allowed the ADI system to evaluate the health of the flexbeam with the second delamination representing a single damage site rather than an additional damage site. This provided nine unique "damage" cases for evaluation by the ADI system. A database was then assembled which included two "no damage" cases and the nine "damage" cases to evaluate the performance of the ADI system in detecting and localizing the damage. These cases are listed in Figure 7 below, along with the output of the ADI system. The "DI" columns present the damage index values for each of the four actuators. The "Damage Detected" column results from comparing the DI values to a threshold of 2.0. The last column provides a rough location estimation based on which actuator had the highest damage index.

| Case # | Site #1 | Site #2 | Baseline | DI #1 | DI #2 | DI #3 | DI #4 | Damage Detected? | Location Identified |
|---|---|---|---|---|---|---|---|---|---|
| 1 | no damage | no damage | no damage | 0.7 | 0.7 | 0.7 | 0.7 | No | - |
| 2 | no damage | no damage | no damage | 1.0 | 1.0 | 0.9 | 0.9 | No | - |
| 3 | small delam | no damage | no damage | 2.8 | 2.7 | 2.3 | 2.4 | Yes | 1 |
| 4 | med. delam | no damage | no damage | 4.1 | 3.9 | 3.6 | 3.6 | Yes | 1 |
| 5 | large delam | no damage | no damage | 19.1 | 13.4 | 12.7 | 11.4 | Yes | 1 |
| 6 | large delam | small delam | no damage | 22.8 | 17.1 | 16.2 | 14.0 | Yes | 1 |
| 7 | large delam | med. delam | no damage | 24.6 | 21.7 | 19.3 | 16.4 | Yes | 1 |
| 8 | large delam | large delam | no damage | 35.2 | 56.4 | 36.6 | 43.9 | Yes | 2 |
| 9 | no damage* | small delam | delam at #1 | 8.6 | 10.4 | 9.7 | 6.8 | Yes | 2 |
| 10 | no damage* | med. delam | delam at #1 | 11.3 | 15.6 | 13.3 | 10.1 | Yes | 2 |
| 11 | no damage* | large delam | delam at #1 | 28.1 | 54.3 | 34.5 | 40.5 | Yes | 2 |

*No damage in reference to the baseline which contains existing damage at site #1

**Figure 7. Tabulated Values of ADI Damage Detection and Localization Results**

The ADI system is able to easily detect the presence of damage for all cases. For this relatively small database, the detection rate achieved is 100% with a zero false alarm rate. The damage index appears to increase as the damage severity is increased, and the location of the damage is consistently identified by the transducer with the largest damage index. A graphical display of the ADI system for an undamaged case (case 2) and a damaged case (case 5) is shown in Figure 8. All damage indexes increase significantly as a result of the damage, but the damage index that increases the most is the transducer which is closest to the damage.

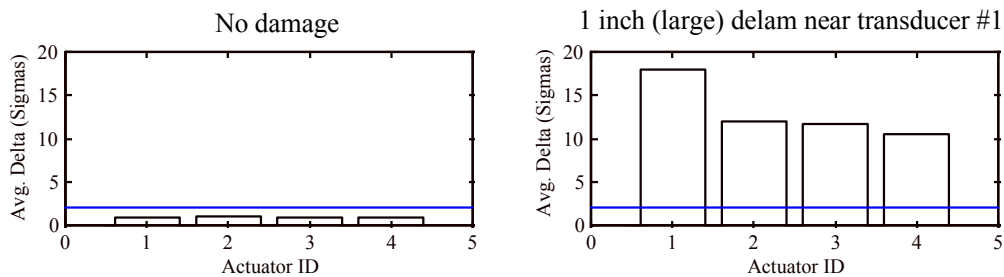

**Figure 8. ADI Results for No Damage (Case 2) and Damage (Case 5) Conditions**

Further processing of these damage index values can result in a more refined localization of the damage. Figure 9 shows the results of one localization algorithm which provides a very accurate estimate of the damage position for the delamination near transducer #1.

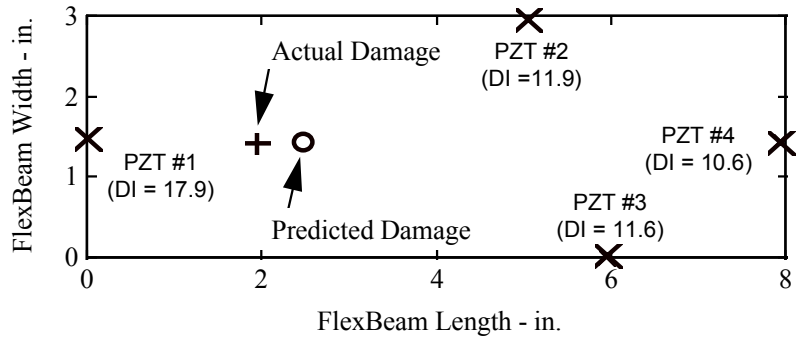

**Figure 9. ADI Localization Results for the One Inch Delamination Near Transducer #1 (Case 5)**

## 4. DISCUSSION AND CONCLUSIONS

The experimental results obtained thus far have demonstrated the potential of using the ADI approach for detecting delaminations in an actual composites structure. While these results are preliminary, they provide the confirmation needed for continuing development of the methodology. Under separate work, the ADI system was shown to also be very effective in detecting simulated fastener failures in a metallic structural test article.[6]

Algorithm development is continuing in order to further refine the ADI system to optimize overall detection, localization, and assessment performance. In addition to the damage index statistics described in this paper, alternate information processing techniques utilizing more spectral features are being investigated. Improvements in data acquisition hardware to allow the usage of higher frequency information are also in work. A distributed architecture involving remote processing nodes is under design to reduce the wiring associated with a network of ADI transducers instrumented on the vehicle. A method to distinguish transducer/bond damage from structural damage has also been developed.

Several issues need to be addressed before field implementation is feasible. These include (1) reducing the size and number of PZT transducers needed to effectively monitor the critical areas of the vehicle structure, (2) evaluating the effect of environmental variation including temperature, humidity, and accumulated load cycles, and (3) conducting extensive damage testing with a large database to quantify overall performance in terms of detection probability, localization accuracy, and false alarm rate. A ground test should be conducted to demonstrate the ADI system with the flexbeam integrated into a complete MD Explorer helicopter rotor system including the hub, pitchcase, and blade. The sensitivity to realistic loading conditions needs to be evaluated. A flight demo will then be needed to fully demonstrate the robustness and effectiveness of the ADI system for health and usage monitoring.

In conclusion, this research has demonstrated that vibration-based structural health monitoring can be accomplished with the Active Damage Interrogation (ADI) approach. The advantages of this approach include simplicity (no need for a model), high sensitivity and accuracy, and low cost implementation. The results obtained thus far indicate considerable promise for integrated structural health monitoring of aerospace vehicles which will enable the practice of condition-based maintenance and consequent reduction in life cycle costs.

## 5. ACKNOWLEDGMENTS

This work was sponsored by the National Science Foundation (NSF) under a Small Business Innovation Research (SBIR) Phase I grant, award number 9660606, teaming NeuroDyne, Inc. and the McDonnell Douglas Corporation (MDC).